\documentclass[a4paper, 10pt]{article}
\usepackage{graphicx}
\usepackage{dcolumn}
\usepackage{bm}
\usepackage{amsmath}
\usepackage{amssymb,amsthm}
\usepackage{epsfig,color}
\usepackage[sans]{dsfont}

\definecolor{nblue}{rgb}{0.2,0.2,0.7}
\definecolor{ngreen}{rgb}{0.2,0.6,0.2}
\definecolor{nred}{rgb}{0.7,0.2,0.2}
\definecolor{nblack}{rgb}{0,0,0}

\newcommand{\tr}{\text{tr}}

\def\tr{\mbox{tr}}

\def\bea{\begin{eqnarray}}
\def\eea{\end{eqnarray}}
\begin{document}

\title{ Discrimination of two-qubit unitaries via local operations and classical communication}

\author{Joonwoo Bae \\ \\[0.5em]
{\it\small  Department of Applied Mathematics, Hanyang University (ERICA), 55 Hanyangdaehak-ro, Ansan, Gyeonggi-do, 426-791, Korea.}}




\maketitle

\begin{abstract}
Distinguishability is a fundamental and operational task generally connected to information applications. In quantum information theory, from the postulates of quantum mechanics it often has an intrinsic limitation, which then dictates and also characterises capabilities of related information tasks. In this work, we consider discrimination between bipartite two-qubit unitary transformations by local operations and classical communication (LOCC) and its relations to entangling capabilities of given unitaries. We show that a pair of entangling unitaries which do not contain local parts, if they are perfectly distinguishable by global operations, can also be perfectly distinguishable by LOCC. There also exist non-entangling unitaries, e.g. local unitaries, that are perfectly discriminated by global operations but not by LOCC. The results show that capabilities of LOCC are strictly restricted than global operations in distinguishing bipartite unitaries for a finite number of repetitions, contrast to discrimination of a pair of bipartite states and also to asymptotic discrimination of unitaries. 
\end{abstract}

\section{Introduction}

Distinguishability is one of the most fundamental measures and, at the same time, a useful tool to characterise capabilities of information applications. In quantum information theory, it has been not only a useful tool to characterise properties of quantum evolution \cite{ref:markov} but also dictate information applications. For instance, the impossibility of perfect discrimination between non-orthogonal quantum states implies that quantum states cannot be perfectly copied \cite{ref:gisin}. Once the no-cloning theorem is applied to a subsystem of entangled states, one can find that entanglement a monogamous correlation that cannot be shared by arbitrarily many parties \cite{ref:werner}. 

Since entangled states are generated by entangling unitary transformations, distinguishability of entangling unitaries may also be related to the way that capabilities of certain quantum information tasks are limited. It, however, has been found that in the aspect of distinguishability, unitary transformations have distinct properties to quantum states. If it is allowed to apply unitaries repeatedly many times, perfect discrimination between two unitaries can be achieved in a finite number of repetitions, whereas quantum states cannot. \cite{ref:acin}. Distinguishability of unitaries can also be improved by ancillary systems, while quantum states are not \cite{ref:italians}. 


In fact, local operations and classical communication (LOCC) play the key role in entanglement theory: those quantum states that can be prepared by LOCC must be separable states, whereas global operations are necessary to create entangled states. Indeed, distinguishability of quantum states has been a useful tool to find a sharp separation between capabilities of LOCC and global operations. A collection of orthogonal bipartite quantum states, which are therefore perfectly distinguishable by global operations, is provided such that they cannot be perfectly distinguishable with LOCC only \cite{ref:nonlocality}. For some cases, however, LOCC are as useful as global operations independently to entanglement present in given quantum states. Given a pair of orthogonal two-qubit states, thus distinguishable by global operations, there is an LOCC protocol that can lead to perfect discrimination between them, no matter how entangled given states are \cite{ref:walgate1} \cite{ref:walgate2}. 

In the case of unitary transformations, distinguishability has been considered in the context of asymptotic discrimination where unitaries can be repeatedly applied. Along the line, results are counterintuitive that any pair of unitaries are perfectly distinguishable in a finite repetitions not only by global operations \cite{ref:acin}, but also by LOCC if they are multipartite unitaries \cite{ref:china1} \cite{ref:china2}. Note that this is independent to entangling capabilities of given unitary transformations. However, this does not directly imply the equivalence between global operations and LOCC. If a number of repetitions is fixed, it is not clear if LOCC reaches distinguishability that global operations would have. Little is known along the line, even in the single-shot scenario that unitaries are applied only once.


We here approach to characterising distinguishability of bipartite unitaries by LOCC in the single-shot scenario and investigate its relation to entangling capabilities. A pair of two-qubit unitaries, both entangling and non-entangling cases, are considered. We show that, on the one hand, any pair of entangling unitaries which do not contain local unitaries (see Eqs. (\ref{eq:dia0}) and (\ref{eq:dia}) for the precise form) are perfectly distinguishable by LOCC whenever they are perfectly distinguishable by global operations. This compares to the single-shot scenario of two-state discrimination where any pair of orthogonal bipartite states, i.e. globally distinguishable, are perfectly distinguishable by LOCC \cite{ref:walgate1} \cite{ref:walgate2}. On the other hand, we show non-entangling unitaries, i.e. local unitaries, that are perfectly distinguishable by global operations but not by LOCC. This contrasts to the asymptotic case where there exists a finite number of repeated applications such that multipartite unitaries are perfectly distinguishable by LOCC \cite{ref:china1} \cite{ref:china2}. Finally, we consider minimum-error discrimination of entangling unitaries and show that LOCC protocols can achieve optimal discrimination by global operations. 

\section{Results}


\subsection{Discrimination of unitary transformations}

Let $U_1$ and $U_2$ denote two unitary transformations we consider throughout. A general framework of distinguishing unitary transformations works as follows \cite{notc}. Suppose that there is a box in which one of two unitaries either $U_1$ or $U_2$ is applied with probabilities $q_1$ and $q_2$ respectively, once a quantum state arrives at the box. After application of unitaries, the resulting state is returned. Let $\rho$ denote the input state to the box, and then the resulting state must be either $U_1 \rho U_{1}^{\dagger} $ or $U_2 \rho U_{2}^{\dagger}$. The optimal discrimination between these states concludes which unitary transformation has been applied in the box. 

For arbitrary two states, minimum-error discrimination has been completely analysed and the success probability is given by, depending on the choice of an input state,
\bea
p_{\mathrm{success}} [\rho]= \frac{1}{2} + \frac{1}{2}  \| q_1 U_1 \rho U_{1}^{\dagger} - q_2 U_2 \rho U_{2}^{\dagger} \|_1 \nonumber
\eea
where $\|\cdot \|_1$ denotes the trace norm, $\| A\| = \tr \sqrt{A^{\dagger}A} $ for hermitian operators $A$. For optimal discrimination between unitaries, an input state should be found such that the success probability is maximised. This introduces optimisation of the distance over input states,
\bea
D (U_1 ,U_2) = \max_{\rho} \|  q_1 U_1 \rho U_{1}^{\dagger} - q_2 U_2 \rho U_{2}^{\dagger}  \|_1 \label{eq:d}
\eea
which we call distinguishability of unitaries. Then, the success probability for unitaries is simplified as $p_{\mathrm{success}} = (1 +D(U_1, U_2))/2 $. 

To compute distinguishability of unitaries, one can in fact restrict the consideration to pure states. This is due to the convexity of distinguishability, as follows. Suppose a pure-state decomposition of an input state $\rho=\sum_i p_i \rho_i$ with $\rho_i = | \psi_i\rangle\langle \psi_i | $. Then, we have 
\bea
D (U_1 ,U_2) & = & \max_{\rho} \|  q_1 \sum_i  p_i U_1 \rho_i  U_{1}^{\dagger} - q_2 \sum_i  p_i U_2 \rho_i  U_{2}^{\dagger}  \|_1  \nonumber\\
& \leq &\max_\rho \sum_{i}  p_i \|  q_1     U_1 \rho_i  U_{1}^{\dagger} - q_2    U_2 \rho_i  U_{2}^{\dagger}  \|_1 \nonumber \\
& \leq & \max_{\rho_i}  \|  q_1     U_1 \rho_i  U_{1}^{\dagger} - q_2    U_2 \rho_i  U_{2}^{\dagger}  \|_1. \nonumber
\eea
Hence, distinguishability of unitaries is obtained by taking pure states as an input to unitary transformations. In addition, the trace distance for pure states has the reciprocal relation as 
\bea
\| q_1|\varphi_1\rangle \langle \varphi_1 | - q_2 |\varphi_2\rangle \langle \varphi_2 |\|_{1}^2=  (1 - 4q_1 q_2 | \langle \varphi_1 | \varphi_2 \rangle|^2 ). \nonumber
\eea
From this, we introduce an equivalent quantity, the fidelity of unitaries $F(U_1, U_2)$, such that
\bea
F(U_1, U_2) = \min_{|\psi\rangle} | \langle  \psi |  U_{1}^{\dagger} U_2 |  \psi \rangle|. \label{eq:dd}
\eea
We then have the following relation between distinguishability and fidelity of unitaries,
\bea
D(U_1, U_2)  = \sqrt{1- 4 q_1 q_2 F^2 (U_1, U_2) }. \nonumber
\eea
This shows the reciprocal relation between fidelity and distinguishability of unitaries. Or, to maximise the success probability, the task is to find state $|\psi\rangle$ that finds the fidelity $F(U_1, U_2)$ of unitaries. 

In fact, discrimination of unitaries with $N$ repetitions \cite{ref:acin} can be rephrased in terms of the fidelity of unitaries. It has considered $N$ repeated applications of unitaries, $U_{1}^{N}$ and $U_{2}^{N}$. It has shown that, if applications of unitaries can be repeated so that the task becomes to discriminating between $U_{1} \times U_1 \times \cdots \times U_1$ and $U_{2} \times U_2 \times \cdots \times U_2$, there exists a finite number of repetitions $N$ and input state $|\psi\rangle$ such that one achieves the case $F(U_{1}^N, U_{2}^{N}) =0$.

\subsection{ Two-qubit unitary transformations and LOCC scenario }
Let us also recall a useful decomposition of two-qubit unitary transformations into entangling and non-entangling parts, and then summarise a discrimination task with LOCC. 

\subsubsection{ Decomposition of unitary transformations}

Two-qubit unitary transformations have a canonical form of decomposition \cite{kraus} \cite{ref:kg}. A two-qubit unitary transformation $W_{AB}$ can be factorised into entangling and local unitaries,
\bea
W_{AB} = (U_A \otimes U_B) W_{AB}[ d ] (V_A \otimes V_B)^{\dagger} \label{eq:dia0} 
\eea
where the entangling part corresponds to the diagonal one $W_{AB}[d]$ and non-entangling ones $\{ U_{A } , U_{B } ,V_{A} ,V_{B}\}$ are local unitaries that only change local basis. The entangling part can be written in a compact way,
\bea
W_{AB}[ d ] = \exp [-i \sigma_{A}^{T} \cdot  d  \cdot \sigma_B] \label{eq:dia}
\eea
with $\sigma_{X} = (\sigma_x , \sigma_y, \sigma_z)$ of Pauli matrices, for $X=A,B$ and a diagonal matrix $d = \mathrm{diag} [v_x,x_y,x_z]$. Note that elements in the diagonal matrix $d$ satisfy the order relation, $\pi /4\geq v_x \geq v_y \geq v_z \geq 0$. The entangling part can be alternatively expressed in its spectral decomposition with Bell states: 
\bea
&& W_{AB} [d]  =   \sum_{j=1}^4  e^{-i \lambda_j } |\Phi_j \rangle  \langle \Phi_j |  ~~\mathrm{where}   \label{eq:decom} \\
&& |\Phi_1\rangle = (|00\rangle + |11 \rangle )/\sqrt{2}, ~~ |\Phi_2\rangle = (|00\rangle - |11\rangle )/\sqrt{2}, ~~\nonumber\\
&&  |\Phi_3\rangle = (|01 \rangle - |10 \rangle )/\sqrt{2}, ~~ |\Phi_4\rangle = (|01\rangle + |10\rangle )/\sqrt{2}. ~~ \nonumber
\eea
Then, these parameters $\{v_{i} \}_{i=x,y,z}$ and $\{\lambda_i \}_{i=1}^4$ are related as follows, 
\bea
\lambda_1 & = & v_x - v_y + v_z \nonumber \\
\lambda_2 & = & - v_x + v_y + v_z \nonumber \\
\lambda_3 & = & - v_x - v_y - v_z \nonumber \\
\lambda_4 & = & v_x + v_y - v_z. \nonumber 
\eea
From the relations and the order among $\{v_{i} \}_{i=x,y,z}$ in the above, it also holds that $\lambda_4 \geq \lambda_1 \geq \lambda_2 \geq \lambda_3$.

The goal is then to find the relation between distinguishability of unitaries and their entangling capabilities. For the purpose, we here restrict our consideration to two entangling unitaries $U_{1,AB}$ and $U_{2,AB}$ which do not contain local parts. That is, both of them are expressed in the diagonal form in Eq. (\ref{eq:dia}). Consequently, their product in Eq. (\ref{eq:dd}) is also in the diagonal form, that is,
\bea
U_{AB} [d] = U_{1,AB}^{\dagger} U_{2,AB} = \exp{ [ -i \sigma_{A}^T \cdot d \cdot \sigma_B ] }   \label{eq:prod}
\eea
for some diagonal matrix $d$. Note that it can also be decomposed into Bell states as it is in Eq. (\ref{eq:decom}). The consideration in Eq. (\ref{eq:prod}) also holds true for a pair of arbitrary two unitaries having their product in the diagonal form. 

\subsubsection{LOCC discrimination}
\label{subsec:locc}

Having specified the form of unitary transformations to be considered, we now introduce how discrimination between a pair of bipartite unitaries works, together with discrimination of bipartite quantum states, by LOCC. Suppose that there are two parties, Alice and Bob, who want to discriminate between two-qubit unitaries denoted by $U_{1,AB}$ and $U_{2,AB}$, are far in distance. Then, there is a box in the middle such that it is not reached by both parties. Once a bipartite state $\rho_{AB}$ comes to the box, one of two-qubit unitaries either $U_{1,AB}$ or $U_{2,AB}$ is applied and then the resulting state, either of the followings
\bea
U_{1,AB}^{\dagger} \rho_{AB} U_{1,AB}~~\mathrm{ or}~~ U_{2,AB}^{\dagger} \rho_{AB} U_{2,AB}\nonumber
\eea
returns to the two parties. Since they are far in distance, LOCC are only available in both stages of preparation of an input state and discrimination between two resulting states. 

As it has been shown in the above, distinguishability of two unitaries is obtained by taking an input state as a pure state, see also Eq. (\ref{eq:dd}). Thus, it suffices for Alice and Bob to prepare product states as follows, 
\bea
|\psi_{AB}\rangle = |\psi_A \rangle \otimes  |\psi_B\rangle. \label{eq:input}
\eea
Then, the next is to discriminate between resulting states, $U_{1,AB} |\psi_{AB} \rangle$ and $U_{2,AB} |\psi_{AB} \rangle$, by LOCC. 

For LOCC discrimination between multipartite states, it has been shown that any pair of orthogonal two-qubit states, i.e., distinguishable by global operations, are also perfectly distinguishable by LOCC \cite{ref:walgate1, ref:walgate2}. An LOCC protocol that perfectly discriminates between orthogonal two-qubit states has also been provided. This means that, as long as the resulting states $U_{1,AB} |\psi_{AB} \rangle$ and $U_{2,AB} |\psi_{AB} \rangle$ are orthogonal, they can be perfectly discriminated by an LOCC protocol. 

Therefore, for two unitaries that are perfectly distinguishable by global operations i.e. $F (U_{1,AB} ,U_{2,AB}) =0$, the LOCC protocol for discrimination of unitaries reduces to finding a product state such that resulting states are orthogonal. That is, from the results on LOCC discrimination \cite{ref:walgate1, ref:walgate2}, we conclude that two-qubit unitaries are perfectly distinguishable by LOCC if there exists a product state $|\psi\rangle$ in Eq. (\ref{eq:input}) such that resulting states $U_{1,AB} |\psi\rangle $ and $U_{2,AB} |\psi\rangle$ are orthogonal, see also Eq. (\ref{eq:prod}): 
\bea
\exists  |\psi_A\rangle, |\psi_B\rangle  ~\mathrm{such~that}~   |   \langle \psi_A |  \langle \psi_B | U_{AB} [ d ] | \psi_A \rangle |\psi_B \rangle | =0. \nonumber
\eea
 
The results for orthogonal bipartite states have been generalised to non-orthogonal states \cite{ref:virmani, ref:chen}. In fact, LOCC protocols can achieve the minimum-error discrimination that is obtained with global operations, regardless of how entangled given a pair of states are \cite{ref:virmani}. Therefore, we can also restrict to our consideration to preparing local states as the input state $|\psi_{AB} \rangle = |\psi_A \rangle\otimes | \psi_B\rangle$ such that resulting states $U_{1,AB} |\psi\rangle $ and $U_{2,AB} |\psi\rangle$ the most distinguishable, that is,
\bea
\min_{|\psi_A\rangle , |\psi_B\rangle} |  \langle \psi_A |  \langle \psi_B | U_{AB} [ d ] | \psi_A \rangle |\psi_B \rangle |, \nonumber
\eea
which is in fact equal to the case that the minimisation is taken with entangled states \cite{ref:virmani}. Then, once Alice and Bob receive the resulting states, they can distinguish them by an LOCC protocol that can achieve optimal discrimination with global operations. 



\subsection{Distinguishability of two-qubit  unitary transformations  }
\label{sec:locc}

We now show discrimination of two-qubit unitary transformations in the following cases: i) when global operations are available, ii) when only LOCC are applied in state preparation and measurement, and iii) when state preparation is performed by LOCC and later global operations are applied in measurement. An input state to unitaries, denoted by $|\psi_{AB}\rangle$, can be written with Bell basis in Eq. (\ref{eq:decom}),
\bea
|\psi_{AB} \rangle = c_1 |\Phi_1\rangle + c_2 |\Phi_2\rangle +c_3 |\Phi_3\rangle +c_4 |\Phi_4\rangle.  \label{eq:con1}
\eea
The state is product if the coefficients satisfy the relations: 
\bea
|\psi_{AB}\rangle = |\psi_A\rangle \otimes |\psi_B\rangle ~\Longleftrightarrow ~c_1^2 + c_3^2 = c_2^2 +c_4^2, \label{eq:con2}
\eea
otherwise, it is entangled.

\subsubsection{Distinguishability by global operations}

\begin{figure}
\begin{center}
\includegraphics[width=9cm]{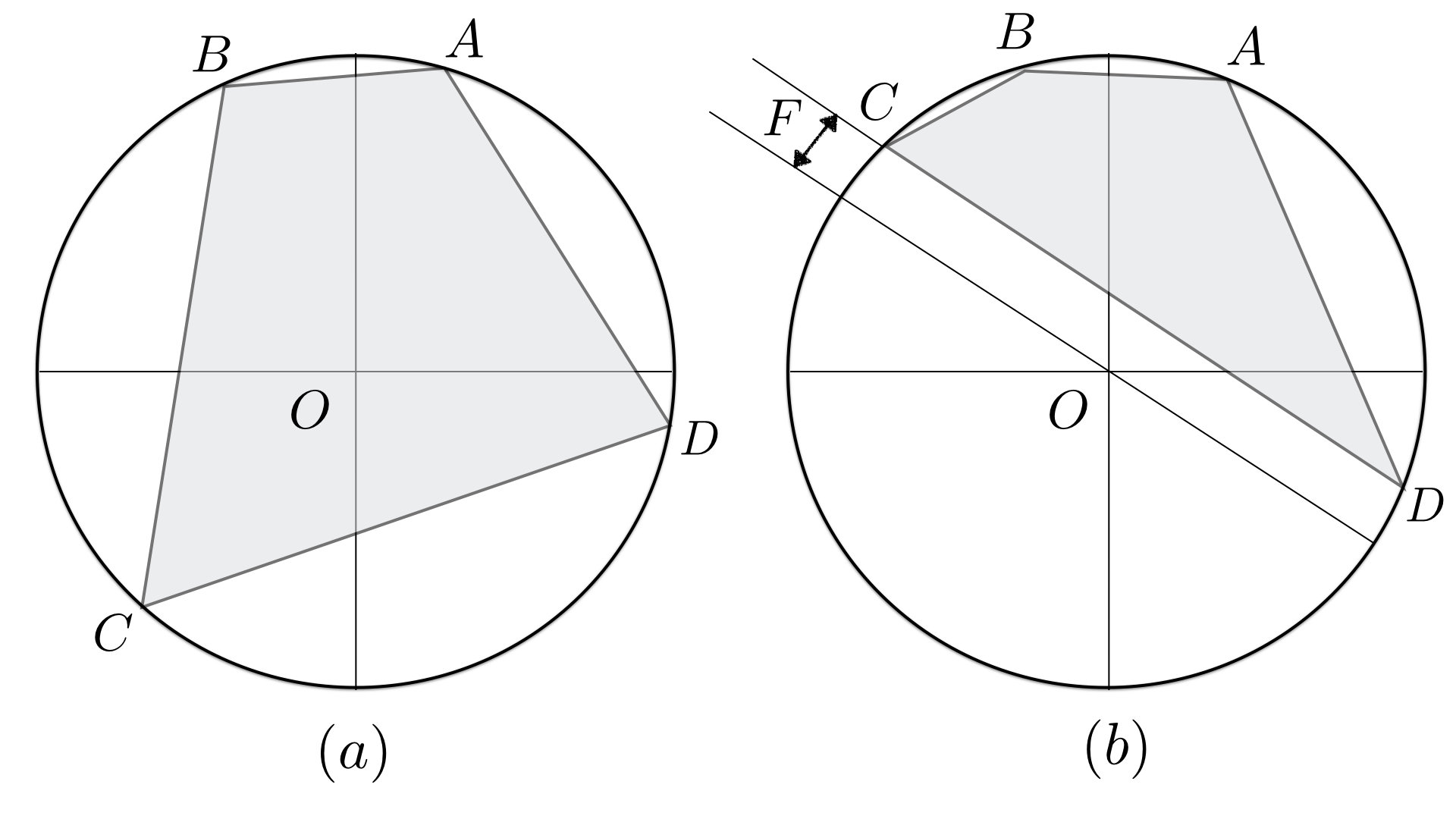}\\
  \caption{ A convex hull in Eq. (\ref{eq:conv}) of the spectrum of a unitary transformation is shown. Vertices $A$, $B$, $C$, and $D$ correspond to $e^{-i\lambda_4}$, $e^{-i\lambda_1}$, $e^{-i\lambda_2}$, and $e^{-i\lambda_3}$, respectively, since $\lambda_4 \geq \lambda_1 \geq \lambda_2 \geq \lambda _3$. Any point in the convex hull can be reached by manipulating input state in Eq. (\ref{eq:con1}). For instance, by taking an input state as Bell states, $|\psi\rangle = |\Phi_j\rangle$ for $j=1,2,3,4$, vertices $A$, $B$, $C$, and $D$ can be reached. In $(a)$ the convex hull contains the origin and there exists an input state such that perfect discrimination is achieved. In $(b)$ the fidelity $F$ corresponds to the distance between the convex hull and the origin, see Eq. (\ref{eq:du}).   }\label{fig1}
  \end{center}
\end{figure}

When global operations are available, Alice and Bob can prepare an arbitrary bipartite state in Eq. (\ref{eq:con1}) such that the fidelity $F(U_{1,AB} , U_{2,AB})$, see Eq. (\ref{eq:dd}), is minimal and distinguishability in Eq. (\ref{eq:d}) is maximised. To be explicit, we have
\bea
F(U_{1,AB} , U_{2,AB}) = \min_{ \sum_j |c_j |^2 =1 } ~  \big|  ~ \sum_{j=1}^4  | c_j |^2 e^{-i \lambda_j }  ~   \big|.   \label{eq:com}
\eea
where $U_{1,AB}^{\dagger} U_{2,AB} $ is an entangling unitary transformation that has a decomposition and parameters in Eq. (\ref{eq:decom}). The goal is then to find an input state i.e. $\{c_j\}_{j=1}^4$ to have Eq. (\ref{eq:com}) minimal. One can observe that the fidelity $F(U_1 , U_2)$ as a convex combination of four complex numbers $\{e^{-i\lambda_j} \}_{j=1}^4$ with weights $ \{ | c_j |^2 \}_{j=1}^4$. Referring to the complex plane, see Figure \ref{fig1}, we find those complex numbers lie in the unit circle and the weights provide a probabilistic mixture of them i.e., $ \sum_j |c_j |^2 =1 $ and $ | c_j |^2 \geq 0$.

Let $\mathrm{conv}( U_{AB})$, where $U_{AB} = U_{1,AB}^{\dagger} U_{2,AB}$, denote the convex hull constructed by $\{e^{-i\lambda_j} \}_{j=1}^4$ with weights $ \{ | c_j |^2 \}_{j=1}^4$ as follows,
\bea
\mathrm{conv }( U_{AB}) =\big\{ \sum_{j=1}^4 |c_j |^2 e^{-i\lambda_j} ~: ~ \sum_{j=1}^4 |c_j |^2 =1  ~\big\}. \label{eq:conv}
\eea
For instance, each one $|c_j |^2 e^{-i\lambda_j} $ corresponds to a vertex of the convex hull. Note that, to construct the convex hull, given is the spectrum $\{ e^{-i \lambda_j } \}_{j=1}^4$ from unitary transformations, and the weights $ \{ | c_j |^2 \}_{j=1}^4$ are manipulated by choosing an input state $|\psi\rangle$ in Eq. (\ref{eq:con1}).

If the convex full contains the origin $O$ in the complex plane, it means that there exist weights $ \{ | c_j |^2 \}_{j=1}^4$ such that $F (U_{1,AB} , U_{2,AB}) =0$, that is, perfect discrimination is achieved. see Figure \ref{fig1}. If the convex hull does not contain the origin, perfect discrimination cannot be achieved and one has to find optimal input state to find optimal discrimination. In this case, distinguishability is then equivalent to the distance between the origin $O$ and the convex hull,
\bea
F(U_{1,AB} , U_{2,AB}) & = &  \mathrm{dist } (\mathrm{conv}(U_{AB} ) , O)  \nonumber\\
& = & \min_{v\in \mathrm{conv (U_{AB} ) }} \| O -  v \|_{2}^2 \label{eq:du}
\eea 
where $\|\cdot \|_2$ denotes the Euclidean norm in the complex plane. 

Finally, it is worth to mention the radical difference between quantum states and unitary transformations in the discrimination scenario. In fact, if unitary transformations can be repeatedly applied, one can always find an input state such that the resulting convex hull contains the origin \cite{ref:acin, ref:italians}. This means the perfect discrimination between two unitaries, which however does not happen in minimum-error discrimination of states.

\subsubsection{ Perfect distinguishability of entangling unitaries: LOCC are as powerful as global operations}

Recall that a pair of two-qubit states that are orthogonal can be perfectly discriminated not only by global operations but also by LOCC \cite{ref:walgate1, ref:walgate2}, where LOCC protocols for the task have been provided. Note also that the result holds true independently to entanglement contained in given two-qubit states. 

Let us now consider a pair of two-qubit unitary transformations that are perfectly distinguishable by global operations, in which two resulting states after application of unitaries are orthogonal. Then, for these unitaries, the problem of distinguishing unitaries reduces to finding an input state prepared by LOCC, i.e. a product state in Eq. (\ref{eq:con2}), such that the resulting states are orthogonal. The cases that resulting states are not orthogonal are also to be discussed. 

Similarly to the convex hull in Eq. (\ref{eq:conv}) introduced with global operations, let $\mathrm{conv_L}$ denote the local convex hull  constrained by the condition in Eq. (\ref{eq:con2}), that is, constructed by state preparation with LOCC: 
\bea
&& \mathrm{conv_L}(U_{AB}) = \mathrm{conv_L}( U_{1,AB}^{\dagger} U_{2,AB} )  \nonumber \\
&& =\big\{ \sum_{j=1}^4 |c_j|^2 e^{-i\lambda_j} ~: ~  \sum_{j=1}^4 |c_j |^2 =1,~c_1^2 + c_3^2 = c_2^2 + c_4^2   ~\big\}. \nonumber
\eea
If the origin is in the local convex hull, $O\in \mathrm{conv_L }(U_{AB})$, it means there exists an input state to unitaries such that the resulting states are orthogonal: hence, two unitaries are perfectly distinguishable by LOCC \cite{ref:walgate1, ref:walgate2}. 

\begin{figure}
\begin{center}
  \includegraphics[width = 9cm]{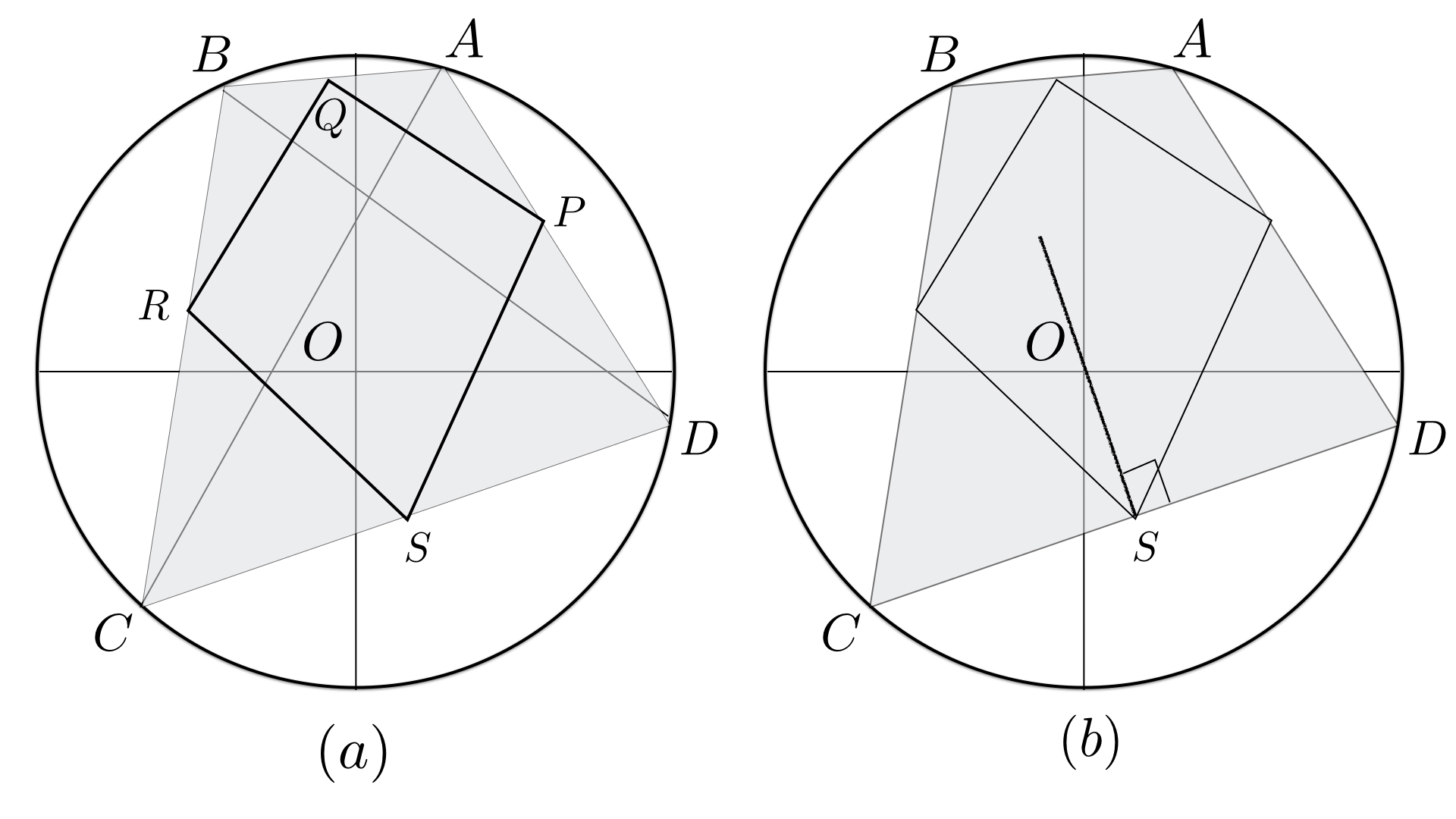}\\
  \caption{ 
A convex hull in Eq. (\ref{eq:conv}) of the spectrum of a unitary transformation is shown. Vertices $A$, $B$, $C$, and $D$ correspond to $e^{-i\lambda_4}$, $e^{-i\lambda_1}$, $e^{-i\lambda_2}$, and $e^{-i\lambda_3}$, respectively, since $\lambda_4 \geq \lambda_1 \geq \lambda_2 \geq \lambda _3$. In (a), the local convex hull is constructed as $\Box PQRS$ where $\{ P,Q,R,S\}$ are midpoints of $\{ DA, AB, BC, CD\}$ respectively. In (b), it is shown that for any of $\{P,Q,R,S \}$ if a line orthogonal to $\{ DA, AB, BC, CD\}$ is drawn, it passes through the origin. It means that if $O\in \mathrm{conv} (U_{AB}) $, then it also holds true that $O\in \mathrm{conv_L}(U_{AB})$.
 }\label{fig2}
\end{center}
\end{figure}

In what follows, we show that a pair of two-qubit unitaries that are perfectly distinguishable, i.e. $O\in \mathrm{conv}(U_{AB})$, there always exists an input state prepared by LOCC such that resulting states are orthogonal, i.e., $O\in \mathrm{conv_L}(U_{AB})$, and thus the two unitaries are perfectly distinguishable by LOCC. We mainly construct the local convex hull $\mathrm{conv_L}$ within a convex hull in Eq. (\ref{eq:conv}), see Figure \ref{fig2}. Let $p_L \in \mathrm{conv_L}(U_{AB})$ denote a point in the local convex hull,
\bea
p_L = \sum_j |c_j|^2 e^{-i\lambda_j}. \label{eq:p}
\eea
From the above, without loss of generality we assume that $\{ c_j\}_{j=1}^4$ are real since $p_L$ only depends on $\{  | c_j |\}_{j=1}^4$. Then, from two conditions of being a product state, $\sum_{i} c_{i}^2 =1$ and $c_1^2 + c_3^2 = c_2^2+c_4^2$, we have $c_1^2 + c_3^2 = c_2^2+c_4^2=1/2$ and introduce parameters $\{ q_i\}_{i=1}^4$:
\bea
\{ q_i = 2 c_i^2\}_{i=1}^4~~\mathrm{so~that}~~ q_1 + q_3 =1,~\mathrm{and}~q_2 + q_4 =1.\nonumber
\eea
With these, a point in the local convex hull can be rewritten as
\bea
&& p_L = \frac{1}{2}[  p_L^{(1,3)}+ p_L^{(2,4)}] \nonumber\\
\mathrm{where} &&p_L^{(1,3)} = (q_1 e^{-i\lambda_1 } +q_3 e^{-i\lambda_3 }),~\mathrm{with}~ q_1+q_3 =1 \nonumber \\
&& p_{L}^{(2,4)} = (q_2 e^{-i\lambda_2 } + q_4 e^{-i\lambda_4 }), ~\mathrm{with}~ q_2 + q_4 =1. \nonumber
\eea
This shows that $p_L$ is found as the midpoint of $p_{L}^{(1,3)}$, a convex combination of $e^{-i\lambda_1}$ and $e^{-i\lambda_3}$, and $p_{L}^{(2,4)}$, a convex combination of $e^{-i\lambda_2}$ and $e^{-i\lambda_4}$. 

In Fig \ref{fig1} (a), the convex hull is given as $\Box ABCD$ where $\{A,B,C,D\}$ correspond to $\{ e^{-i\lambda_4}, e^{-i\lambda_1}, e^{-i\lambda_2}, e^{-i\lambda_3} \}$, respectively. Then, as it is shown in the above, the local convex hull is constructed as the set of midpoints of $p_{L}^{(1,3)}$, convex combinations of $B$ and $D$, and $p_{L}^{(2,4)}$, convex combinations of $A$ and $C$. The local convex hull thus corresponds to $\Box PQRS$ where $\{P,Q,R,S\}$ are midpoints of $\{ DA, AB, BC, CD\}$, respectively. This can be constructed as follows, see also Figure \ref{fig2} (a). First, note that $p_{L}^{(1,3)}$ corresponds to any point on the line $BD$ and $p_{L}^{(2,4)}$ on the line $AC$. Then, local convex hull is the collection of all midpoints of points on $AC$ and $BD$. For instance, taking one point $A$ and points on $BD$, one can find the midpoints on $PQ$. Or, taking $C$ and $BD$, midpoints $RS$ are found. In this way, one can see that $\Box PQRS$ is the local convex hull. 

Finally, having constructed the local convex hull within the convex hull, it remains to show that if the origin $O$ is in the convex hull, then it is also in the local convex hull. This follows from geometric properties of a circle: if a line is drawn within a circle where both end points touch the circle, then another line from the midpoint such that it is orthogonal to the original line passes through the origin. In this case, see Fig \ref{fig2} (b), let us consider a line $CD$ and $S$ its midpoint. Applying the property here, it holds true that if a line is drawn from $S$ such that it is orthogonal to $CD$ then it passes the origin. This proves that if the origin is in the convex hull, it is also in the local convex hull. Thus, we have shown that a pair of entangling unitaries that are perfectly distinguishable by global operations can also be perfectly discriminated by LOCC.


\subsubsection{Distinguishability of non-entangling unitaries: global operations are strictly more powerful than LOCC}

While it has been shown so far that distinguishing entangling unitaries LOCC are as powerful as global operations, we here show that it cannot be generalised to arbitrary unitaries. We provide a pair of non-entangling unitaries, i.e. local unitary transformations, that are perfectly distinguishable by global operations but not by LOCC. 

Let us consider two local unitaries, the product of which is 
\bea
U = U_{A}\otimes U_{B},~~\mathrm{with} ~~U_{A} = U_{B} = |0\rangle\langle0| + e^{i\pi/2}|1\rangle \langle 1|. \label{eq:lu} 
\eea 
The convex hull of $U_A\otimes U_B$ is the triangle constructed with three vertices $\{0,e^{i\pi}, e^{i\pi/2} \}$ containing the origin. Thus, two unitaries in this case are perfectly distinguishable by global operations. For instance, by taking input state $|\psi\rangle = (|00\rangle +  |11\rangle)\sqrt{2}$ two unitaries are perfectly distinguishable, $\langle \psi |  U  |\psi\rangle = 0$. However, for a product state $|\psi\rangle = |\psi_A\rangle \otimes |\psi_B\rangle$, we have \bea
\langle \psi| U_A\otimes U_B | \psi\rangle = | \langle  \psi_A |U_A | \psi_A \rangle  |^2     | \langle  \psi_B |U_B | \psi_B\rangle  |^2. \nonumber 
\eea
This shows that, unless either of the products $U_A$ or $U_B$ is perfectly distinguishable, it is not possible to perfectly discriminate between a pair of local unitaries having product $U_A \otimes U_B$. More precisely, the convex hull of $U_A$ does not contain the origin, neither does the convex hull of $U_B$. Hence, they cannot be perfectly discriminated by LOCC.

\subsubsection{ Optimal distinguishability of entangling unitaries: LOCC are as powerful as global operations}

Coming back to entangling unitaries having their product in the diagonal form in Eq. (\ref{eq:prod}), suppose that for any input state $|\psi\rangle$, the resulting states $U_{1,AB} |\psi \rangle$ and $U_{2,AB} |\psi\rangle$ are not orthogonal. In this case, the alternative is to find an optimal state such that the resulting states are most distinguishable. And then, if global operations are available, one may apply optimal state discrimination between them \cite{ref:hel}. Remarkably, the results for orthogonal states \cite{ref:walgate1, ref:walgate2} have been generalized to non-orthogonal ones \cite{ref:virmani, ref:chen}. Namely, for a pair of non-orthogonal multipartite states, one can always find an LOCC protocol that achieves the minimum-error discrimination by global operations. Therefore, global operations in optimal discrimination of non-orthogonal states can be replaced by LOCC protocols. 

Then, what remains for optimal discrimination of unitaries is to compare state preparation, that is, local states and entangled states as input states to unitaries. This means that one has to find the convex hull in Eq. (\ref{eq:du}) with local states. That is, denoted by $F_L$ the fidelity from the local convex hull, we have the following as the fidelity between unitaries,
\bea
F_L(U_{1,AB}, U_{2,AB}) & = &  \mathrm{dist} (\mathrm{conv_L} (U_{AB}) ,0 ) \nonumber \\
& = & \min_{v \in \mathrm{conv_L}(U_{AB}) } {\|  O-v \|_2^2 } \nonumber
\eea
with $\| \cdot \|_2$ denotes the Euclidean norm in the complex plane. One can then compare fidelities, $F$ in Eq. (\ref{eq:du}) and $F_L$ in the above.

\begin{figure}
\begin{center}
  \includegraphics[width = 9cm]{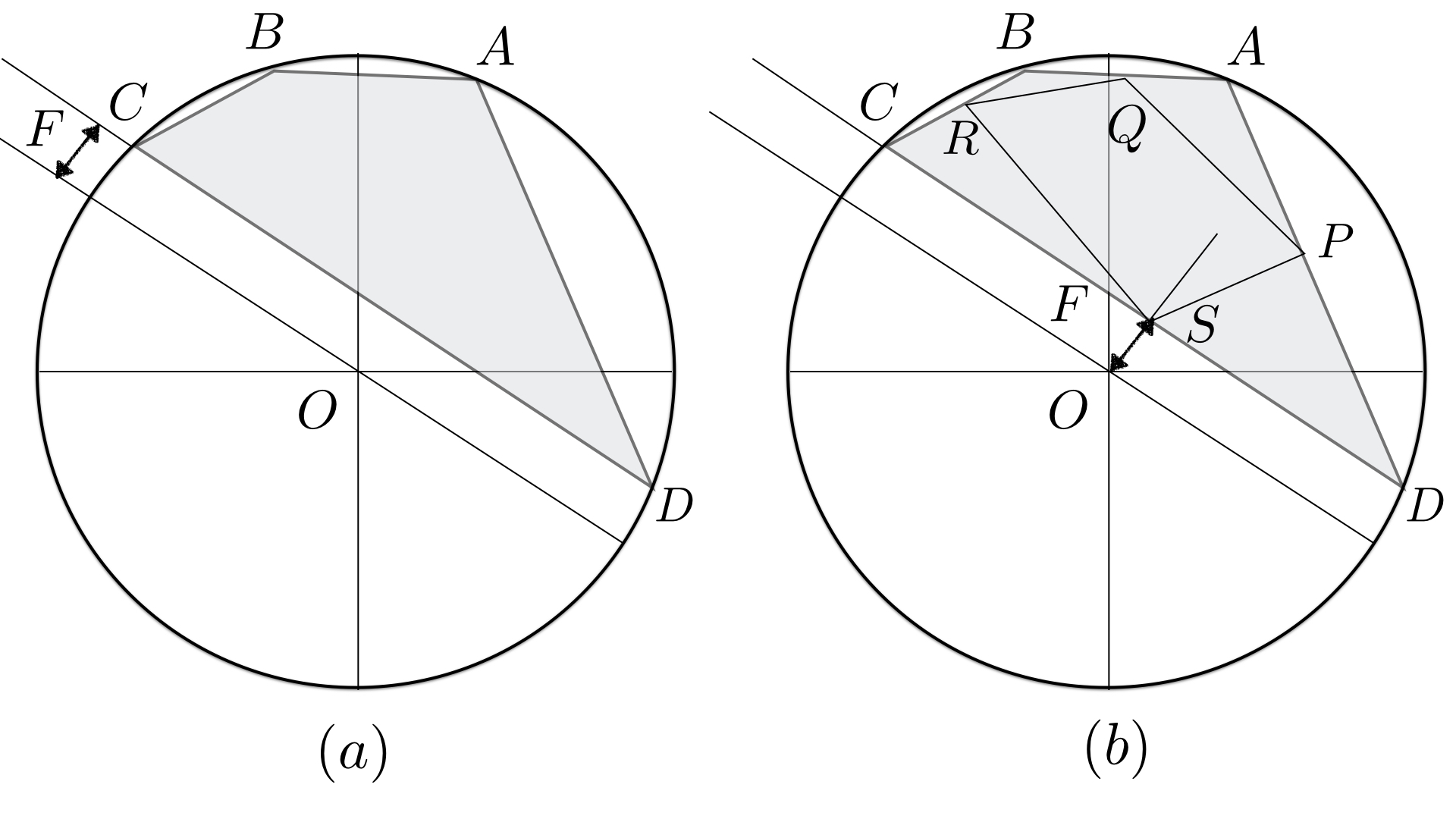}\\
  \caption{ Suppose that resulting two states $U_1 |\psi\rangle$ and $U_2 |\psi\rangle$ are not orthogonal and do not lead to perfect discrimination between two unitary transformations. (a) The fidelity is given by the distance between origin and the convex hull, see also Eq. (\ref{eq:du}). (b) The local convex hull constructed by a product state is found as $\Box PQRS$ where $\{P,Q,R,S \}$ are midpoints of $\{DA,AB,BC,CD \}$, respectively. Note that, while an input state is prepared locally, discrimination between resulting states is performed by global operations. Then, the distance between the local convex hull and the origin corresponds to the distance $OS$, since $OS$ is orthogonal to $DC$ and gives the minimal distance between the local convex hull and the origin. 
 }\label{fig3}
\end{center}
\end{figure}

We recall that in the complex plane in Figure \ref{fig2}, the local convex hull is constructed by connecting midpoints. In Figure \ref{fig3}, when the convex hull does not contain the origin, the local convex hull is constructed in the same way as $\Box PQRS$ where $S$ corresponds to the midpoint of $DC$. Moreover, $OS$ is orthogonal to $DC$ and thus gives the minimal distance between the local convex hull to the origin. In fact, this shows that the minimal distance is equal to the distance between the convex hull $\Box ABCD$ and the orgin, and hence, we have
\bea
F(U_{1,AB}, U_{2,AB}) = F_L(U_{1,AB}, U_{2,AB}). \nonumber
\eea
We have shown that, for distinguishing entangling unitaries having their product in the diagonal form in Eq. (\ref{eq:prod}), there exists an LOCC protocol of state preparation and measurement that achieves optimal discrimination by global operations.

\section{Discussion}

We have considered distinguishability of bipartite two-qubit unitary transformations in the following cases:  i) global operations are available, ii) LOCC are only available, and iii) LOCC are available in state preparation and global operations can be applied in measurement for state discrimination. We have then compared the three cases. We first recall that for a pair of bipartite quantum states, if they are orthogonal, they are perfectly distinguishable not only by global operations but also LOCC \cite{ref:walgate1, ref:walgate2}. The capability of LOCC, however, does not generalise to bipartite unitaries. We have shown a pair of unitaries having their product in the form in Eq. (\ref{eq:lu}) which are perfectly distinguishable by global operation operations but not by LOCC.  
We also recall the result of asymptotic discrimination of unitaries: any pair of unitaries can be perfectly discriminated in finite repetitions of unitaries by global operations \cite{ref:acin}, and also by LOCC for multipartite unitaries \cite{ref:china1, ref:china2}. We here have shown that the capability of LOCC cannot apply to cases when the number of repetitions is fixed: if unitary transformations can be applied only once, there exist local unitaries that are perfectly distinguishable by global operations but not by LOCC. 

Finally, we have shown that a pair of entangling unitaries having their product in Eq. (\ref{eq:prod}) that contains no local unitaries, if they are distinguishable by global operations, can also be perfectly distinguishable by LOCC. For cases that they are not perfectly distinguishable by global operations, we have shown that there exists an LOCC protocol of state preparation and measurement that can reach the capability of global operations. Our results have found that relations between entangling capabilities and distinguishability of unitaries are highly non-trivial, while entanglement contained in states is closely related to distinguishability and related tasks, e.g. \cite{ref:bae}. We envisage deeper relations between distinguishability and entangling capabilities of unitary transformations in characterisations of capabilities of quantum information tasks.

\section*{Acknowledgements }

We thank A. Ac\'in, U. Sen, A. Sen, and G. Chiribella for helpful discussions. This work is supported by Institute for Information \& communications Technology Promotion(IITP) grant funded by the Korea government(MSIP) (No.R0190-15-2028, PSQKD).








\begin{thebibliography}{10}
\bibitem{notc}
Note that, in order to compare to the results in Refs. \cite{ref:china1} \cite{ref:china2}, we here do not consider ancillary systems in discrimination between unitaries.
\end{thebibliography}
\end{document}